\documentclass[preprint]{iucr}              % DO NOT DELETE THIS LINE

     \journalcode{J}              % Indicate the journal to which submitted
\usepackage{amsmath, amsthm, amssymb, mathrsfs, url}
\usepackage{graphicx}
\usepackage{relsize}
\usepackage[margin=1.0in]{geometry}
\usepackage{color}
\usepackage{multirow}
\usepackage{dcolumn}% Align table columns on decimal point
\graphicspath{{./figs/}}

\def\Rw{$R_w$}

\begin{document}                  % DO NOT DELETE THIS LINE

     %-------------------------------------------------------------------------
     % The introductory (header) part of the paper
     %-------------------------------------------------------------------------

     % The title of the paper. Use \shorttitle to indicate an abbreviated title
     % for use in running heads (you will need to uncomment it).

\title{Symmetry-mode analysis for local structure investigations using pair distribution function data}
\shorttitle{Symmetry-driven PDF analysis}

     % Authors' names and addresses. Use \cauthor for the main (contact) author.
     % Use \author for all other authors. Use \aff for authors' affiliations.
     % Use lower-case letters in square brackets to link authors to their
     % affiliations; if there is only one affiliation address, remove the [a].

\author[a]{Parker~K.}{Hamilton}
\author[b,c]{Jaime~M.}{Moya}
\author[d,e]{Alannah~M.}{Hallas}
\author[c,f]{E.}{Morosan}
\author[a]{Raju}{Baral}
\cauthor[a]{Benjamin~A.}{Frandsen}{benfrandsen@byu.edu}{}

\aff[a]{Department of Physics and Astronomy, Brigham Young University, Provo, Utah 84602, \country{USA}}
\aff[b]{Applied Physics Program, Rice University, Houston, Texas 77005, \country{USA}}
\aff[c]{Department of Physics and Astronomy, Rice University, Houston, Texas 77005, \country{USA}}
\aff[d]{Quantum Matter Institute, University of British Columbia, Vancouver, BC, V6T 1Z4, \country{Canada}}
\aff[e]{Department of Physics \& Astronomy, University of British Columbia, Vancouver, BC, V6T 1Z1, \country{Canada}}
\aff[f]{Rice Center for Quantum Materials, Rice University, Houston, Texas 77005, \country{USA}}

     % Use \shortauthor to indicate an abbreviated author list for use in
     % running heads (you will need to uncomment it).

%\shortauthor{Soape, Author and Doe}

     % Use \vita if required to give biographical details (for authors of
     % invited review papers only). Uncomment it.

%\vita{Author's biography}

     % Keywords (required for Journal of Synchrotron Radiation only)
     % Use the \keyword macro for each word or phrase, e.g. 
     % \keyword{X-ray diffraction}\keyword{muscle}

%\keyword{keyword}

     % PDB and NDB reference codes for structures referenced in the article and
     % deposited with the Protein Data Bank and Nucleic Acids Database (Acta
     % Crystallographica Section D). Repeat for each separate structure e.g
     % \PDBref[dethiobiotin synthetase]{1byi} \NDBref[d(G$_4$CGC$_4$)]{ad0002}

%\PDBref[optional name]{refcode}
%\NDBref[optional name]{refcode}

\maketitle                        % DO NOT DELETE THIS LINE

\begin{synopsis}
We demonstrate the use of symmetry-mode analysis to detect and characterize structural distortions from pair distribution function data by refining mode amplitudes directly. This approach is successfully applied to a subtle but long-range distortion in TiSe$_2$, as well as a large but highly localized distortion in MnTe. We also introduce the open-source python packages \texttt{isopydistort} and \texttt{isopytools} used to carry out this analysis within the open-source DiffPy framework.
\end{synopsis}

\begin{abstract}
Symmetry-adapted distortion modes provide a natural way to describe distorted structures derived from higher-symmetry parent phases. Structural refinements using symmetry-mode amplitudes as fit variables have been used for at least 10 years in Rietveld refinements of the average crystal structure from diffraction data; more recently, this approach has also been used for investigations of the local structure using real-space pair distribution function (PDF) data. Here, we further demonstrate the value of performing symmetry-mode fits to PDF data through the successful application of this method to two topical materials: TiSe$_2$, where we detect the subtle but long-range structural distortion driven by the formation of a charge density wave, and MnTe, where we characterize a large but highly localized structural distortion in terms of symmetry-lowering displacements of the Te atoms. The analysis is performed using fully open-source code within the DiffPy framework using two packages we developed for this work: \texttt{isopydistort}, which provides a scriptable interface to the ISODISTORT web application for group theoretical calculations, and \texttt{isopytools}, which converts the ISODISTORT output into a DiffPy-compatible format for subsequent fitting and analysis. These developments expand the potential impact of symmetry-adapted PDF analysis by enabling high throughput analysis and removing the need for any commercial software.
\end{abstract}

     %-------------------------------------------------------------------------
     % The main body of the paper
     %-------------------------------------------------------------------------
     % Now enter the text of the document in multiple \section's, \subsection's
     % and \subsubsection's as required.

\section{Introduction}

Total scattering techniques are a powerful tool for understanding the structure of solids \cite{billi;cc04,keen;crev20}. These methods include in the analysis not only the Bragg scattering that arises from the long-range correlations of the average structure, but also the diffuse scattering that originates from short-range correlations deviating from the average structure. The use of total scattering therefore gives a window into short-range structural correlations that are missed by Bragg peak analysis and other probes of the average structure.

Pair distribution function (PDF) analysis is a well-established analytic technique for total scattering data that involves Fourier transforming the total scattering pattern to yield a map of the interatomic distances as a function of real-space distance $r$~\cite{egami;b;utbp12}. The scattering arising from correlated distortions of the local structure is thereby recast from broad, diffuse signals in reciprocal space to sharper, more localized signals in real space. This can lend a more intuitive interpretive power to the real-space PDF data and also allow for more effective modeling of the local structure~\cite{billi;s07}.

% Unnecessary text, saving in case it becomes useful at some later point.
%One common approach to modeling PDF data is ``small-box modeling'', in which structural parameters corresponding to a periodic unit cell are optimized to provide the best fit to the PDF data over the selected range of $r$. These structural parameters typically include the unit cell dimensions, atomic positions, and atomic displacement parameters (ADPs), similar to what would be optimized in a conventional Rietveld refinement using Bragg diffraction data. However, with PDF data, one has the additional flexibility of selecting a desired range of $r$ for any given fit, which determines the length scale over which the refined parameters are representative. A fitting range of $r<10$~\AA, for example, provides information only about the sub-nanometer local structure, whereas a fitting range of, e.g., $40 < r < 60$~\AA\, provides information on this longer length scale without any influence from shorter-range correlations. Thus, PDF analysis provides information about the length scale of various structural features that cannot be accessed by conventional Rietveld refinements using diffraction data.

A typical objective of PDF analysis is to identify and characterize symmetry-breaking distortions in the local structure that average to zero over longer length scales. Local symmetry breaking of this type can provide crucial information about the underlying physics of the material under investigation~\cite{billi;cc04,billi;s07,young;jmc11,keen;n15,dagot;s05,tokur;np17,desgr;cej18,zhu;advs21}. In the commonly used ``small-box modeling'' or ``real-space Rietveld'' approach, one often starts with a structural model appropriate for the average structure, fits the model to the low-$r$ portion of the data, and looks for systematic misfits that provide evidence for and information about any short-range structural distortions. The next step is then typically to test lower-symmetry models against the data to see if the fit improves, and if so, attempt to determine what physical insights can be gained from the lower-symmetry model~\cite{chepk;acso22}.

Two commonly occurring obstacles can make this strategy difficult. First, selecting candidate lower-symmetry models can often be an \textit{ad hoc} process requiring tedious guesswork by the PDF analyst, and even if some distorted model that improves the fit is identified, it may not be clear whether other distorted models could provide equally good or even better fits. In some cases, theoretical input or experimental data from other probes such as high-resolution diffraction or Raman spectroscopy~\cite{forse;cm15} may help guide the choice of distorted model, but such input is not always available. Second, lower-symmetry models will naturally have a greater number of free parameters (e.g. atomic positions and atomic displacement parameters [ADPs]) to be optimized in the fit, which can impede the optimization routine from converging reliably.

These difficulties apply both to PDF and Rietveld analysis of diffraction data. A promising way to overcome these challenges is to use a symmetry-adapted fitting scheme using the tools of group theory, as has been demonstrated for Rietveld refinements with diffraction data~\cite{kerma;aca12,gawry;prb19}. Using software such as ISODISTORT~\cite{campb;jac06,stoke;webIso}, one can start with a parent structure and systematically explore distorted child structures down to arbitrarily low symmetry and up to an arbitrarily large supercell. Comparing the fit results allows one to build a more complete picture of potential distorted structures that could explain the data. This addresses the first of the previously mentioned challenges. In addition, one can define as fit variables the amplitudes of the symmetry modes allowed for a given structure, in place of the atomic coordinates for individual atoms that are conventionally used as fit variables. A single variable in the form of a mode amplitude may therefore control the collective motion of several distinct atoms. Because structural distortions often correspond to a small number of symmetry modes taking nonzero amplitudes~\cite{kerma;aca12}, the use of symmetry mode amplitudes as a basis for the fit variables can greatly reduce the number of variables required to achieve a good fit, leading to more robust convergence. In addition, the symmetry modes relevant for a given structural distortion can be identified directly from the fit, providing greater clarity to the physical interpretation of the fit results. This feature of a symmetry-adapted fitting scheme addresses the second of the challenges mentioned previously.

Recently, symmetry-mode analysis has also been applied to PDF data~\cite{bird;jac21}. Bird \textit{et al.} introduced a procedure for refining distortion mode amplitudes from PDF data that entails (1) downloading the symmetry mode information from the ISODISTORT web application, (2) running a script to convert the output into a format that can be read by the commercial structural refinement software TOPAS-Academic~\cite{coelh;jac18}, and (3) iterating through a series of refinements in TOPAS to identify and evaluate candidate distortion modes. The approach was successfully used to characterize dynamic distortions in the negative thermal expansion material ScF$_3$ and short-range distortions in the prototypical ferroelectric BaTiO$_3$, demonstrating that a symmetry-adapted fitting scheme can be applied to PDF analysis, providing the aforementioned advantages over conventional fitting methods. It has since been applied to ReO$_3$~\cite{bird;prb21} and BaTiO$_3$ under pressure~\cite{herli;prb22}, suggesting that this approach is widely applicable.

% Unnecessary text, saving in case it is useful at some later point
%\color{black} A challenge in understanding a short or long range structural distortion is fitting the distorted atomic positions. The issues is made worse when a distortion of a longer period of than the primitive lattice is present, requiring the analysis of a super cell, increasing the number of free parameters. We have implemented a method to effectively reduce the parameter space for fitting distorted structures by considering symmetry constrained modes of atomic distortion instead of the usual Cartesian coordinate approach. 

%One common method of identifying the presence of local structural distortions is to fit a model of the average structure (e.g. as determined by conventional Rietveld refinement of the diffraction pattern) to a given range of PDF data and then inspect the fit residual. A distortion that is present in the local structure but ................not the average structure will manifest in systematic misfits at low $r$ that disappear as $r$ increases. 

Here, we further demonstrate the value of symmetry-adapted PDF analysis and introduce alternative software tools for implementing this approach in an automatable, fully open-source way. We have developed the open-source python packages \texttt{isopydistort} and \texttt{isopytools}, which respectively enable automatic calls to ISODISTORT from a python script and convert the ISODISTORT output into a format suitable for structural refinements using the open-source DiffPy library~\cite{juhas;aca15}. This work broadens the impact of symmetry-adapted PDF analysis by eliminating exclusive reliance on commercial software and making possible high throughput analysis. We use this method to study two complementary examples drawn from materials of significant recent interest: the charge-density-wave (CDW) compound TiSe$_2$, which displays a long-range but subtle structural distortion, and the high-performance thermoelectric candidate MnTe, where we observe a large but highly localized distortion. In both cases, we identify the relevant distortions and achieve excellent PDF fits through a systematic and automated exploration of all possible symmetry modes. The results further emphasize the advantages of using symmetry modes when performing structural refinements and suggest that the open-source tools introduced here can be successfully applied to a wide variety of materials.

\section{Methodology: Refining distortion models using symmetry modes}

\begin{figure}
    \centering
    \includegraphics[width=12.0cm]{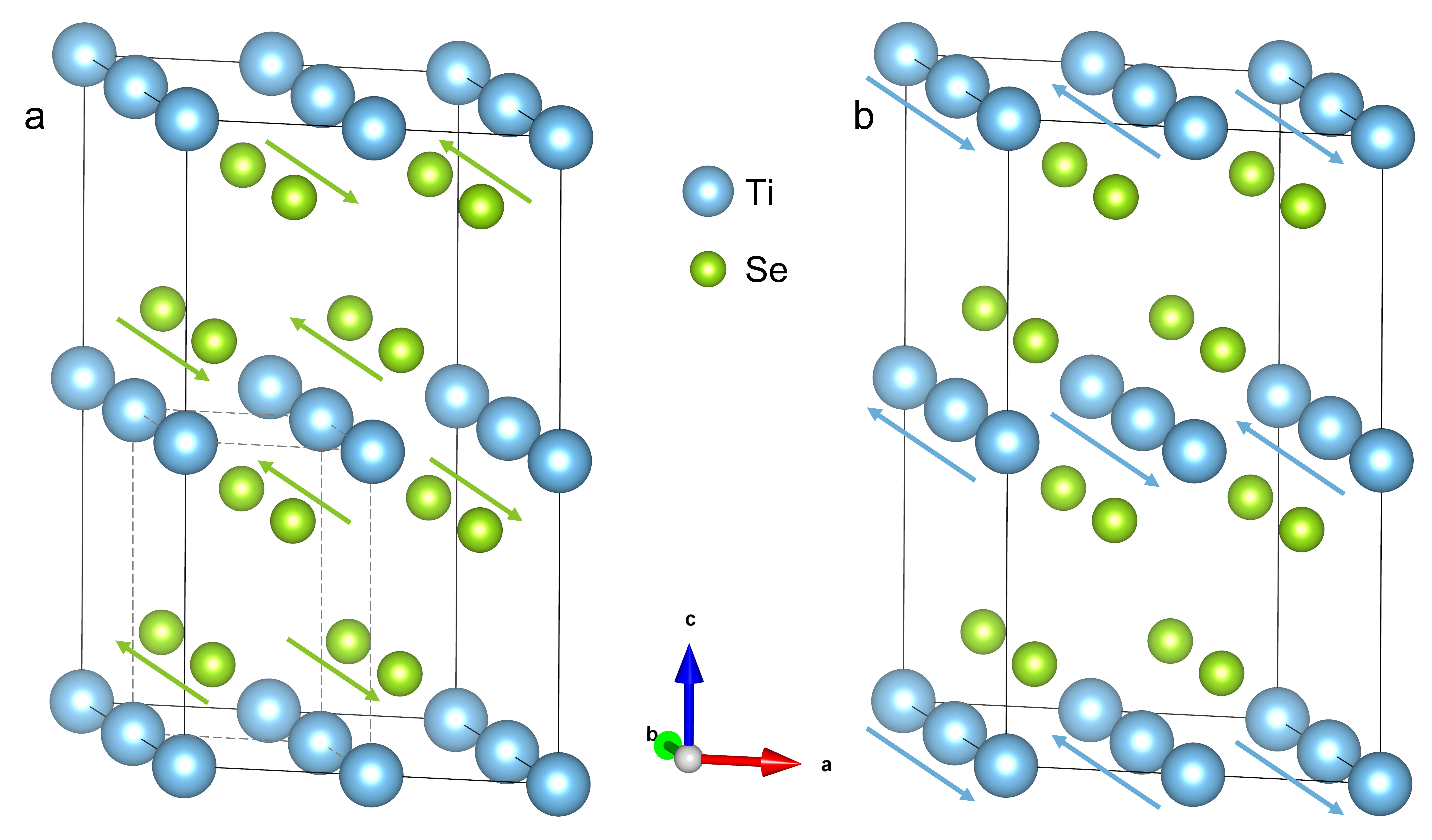}
    \caption{A $2 \times 2 \times 2$ supercell of the $P\overline{3}m1$ structure of TiSe$_2$, visualized using VESTA~\cite{momma;jac11}. The arrows depict distortions belonging to the $L_1^-$ irrep. (a) The Se distortion corresponding to the $E(c)$ mode. The dashed lines indicate the parent cell. (b) The Ti distortion corresponding to the $E_u(c)$ mode.}
    \label{fig:sym_mode}
\end{figure}

A distortion removes certain symmetry elements of the parent space group. The remaining symmetry elements that persist in the distorted structure comprise the ``distortion symmetry'' or  ``isotropy subgroup'' and form the space group of the distorted structure. In this work, we consider only displacive distortions involving changes in the positions of the atoms. To test distorted structural models against PDF data, we use ISODISTORT, which can generate all possible distortion symmetries of a given parent structure for a user-specified supercell and space-group type. Furthermore, ISODISTORT can express the distortion models using either the traditional $xyz$ basis of atomic coordinates or the symmetry-informed basis of symmetry-mode amplitudes, which are constructed from the basis vectors of the irreducible representation (irrep) associated with the given distortion. These two bases---the $xyz$ basis and the mode-amplitude basis---are related by a linear transformation calculated by ISODISTORT. In general, a single mode amplitude can affect the positions of multiple atoms, and the displacement of a single atom may be determined by the amplitudes of multiple modes. Since both bases describe the same distortion space, the number of symmetry modes always equals the number of independent atomic coordinates. For $P$1 symmetry, there are no constraints on atomic positions, so for a supercell with $N$ atoms, there are 3$N$ atomic coordinates and 3$N$ symmetry modes. For higher symmetries, of course, there will be fewer degrees of freedom, since not all atomic displacements will be independent. An example of a symmetry mode from the TiSe$_2$ study described later is shown in Fig. \ref{fig:sym_mode}.

The traditional approach to refining a distorted structural model is to use the $xyz$ basis as fit variables and refine the positions of the atoms directly. However, if the correct distortion symmetry is not known (and sometimes even if it is), there may be so many degrees of freedom that standard fitting algorithms have difficulty converging to a good result. Recognizing that distortions are often dominated by one or a few modes while the others remain inactive, it can be much more effective to use mode amplitudes as fit variables rather than atomic coordinates \cite{kerma;aca12}. A small number of free parameters corresponding to the mode amplitudes can therefore capture collective displacements of multiple atoms that would otherwise require a larger number of atomic coordinates to describe. Furthermore, the fitting scheme can be organized such that one symmetry mode is tested at a time while all the others remain fixed, which allows the identification of relevant modes much more effectively than would be possible by cycling through the individual atomic coordinates. Finally, this symmetry-informed approach directly reveals which symmetry modes are active in a given structure, yielding insight into the underlying physics at play in the material.

%This approach has been applied to Rietveld refinement \cite{kerma;aca12,gawry;prb19} and PDF analysis \cite{bird;jac21} using the \textit{TOPAS} commercial software suite. In these works, candidate distortions of a parent structure were generated in ISODISTORT by selecting a supercell size (which could be the same size as the parent unit cell) and lowering the assumed symmetry of the distorted structure to $P$1, which therefore generates all possible distortion modes. The mode amplitudes were then refined individually in TOPAS, with some threshold fit improvement used to distinguish active from inactive modes. An additional refinement with all active modes included simultaneously can then be performed, producing the final distorted structure.   

% RESOLVED
% We need to add a few sentences here describing the new contributions of our work.
% I just re-read the Bird paper for SAPA and their code is literally just a IO translator to plug ISODISTORT outputs into TOPAS. They also only you to turn whole irrep sets off and on together, which seems a bit shortsighted.

In this work, we use original open-source python packages \texttt{isopydistort}, which interfaces directly with the ISODISTORT server online through a python script inspired by similar functionality available in GSAS-II~\cite{toby;jac13}, and \texttt{isopytools}, which converts the ISODISTORT output into a format compatible with the open-source DiffPy suite for PDF fitting and analysis. These developments remove the need to click through the  ISODISTORT web tool to generate the distortion models, providing a significantly greater degree of automation and flexibility when performing symmetry-mode fits to PDF data, all within a fully open-source software environment. Below we describe the typical workflow, which can be carried out completely within a python script or jupyter notebook using the functions defined in our python packages.
\begin{enumerate}
\item Upload a crystallographic information file (CIF) corresponding to the parent structure (e.g. as determined by Rietveld refinements to diffraction data or by PDF fits conducted over a long fitting range) to the ISODISTORT server, together with an optional dictionary of ISODISTORT arguments such as the lattice basis of the supercell relative to the parent cell or the desired space group of the child structure ($P1$ by default). This is done with the \texttt{isopydistort} package. The code interfacing with ISODISTORT then automatically works through the steps to produce a human-readable file (in TOPAS format) containing all the symmetry modes compatible with the specified supercell and space group. This file is automatically downloaded from the ISODISTORT server. Other ISODISTORT output files can also be requested by the user, such as a CIF or interactive distortion visualization file.
\item Create a DiffPy-compatible structure object from the ISODISTORT output using the \texttt{isopytools} package.
\item Input the symmetry-mode amplitudes as variables to be optimized in the fit, together with the usual fit variables such as a scale parameter, lattice parameters, and ADPs. The \texttt{isopytools} package automatically implements the appropriate constraints relating the mode amplitudes to the atomic coordinates based on the ISODISTORT output.
\item Use \texttt{diffpy.srfit} to carry out the fit however the user desires. For example, the user can cycle through each mode amplitude individually, refine multiple amplitudes in arbitrarily defined groups, perform multiple fits with random starting values, impose a cost for activating additional modes~\cite{kerma;aca12}, establish a threshold for improvement in the fit quality to consider a mode active~\cite{bird;jac21}, etc.
\item Modify the \texttt{diffpy.srfit} script as desired for automatic batch fits to multiple data sets, different fitting ranges, etc.
\end{enumerate}
We followed this basic workflow to study distortions present in TiSe$_2$ and MnTe.

\section{Experimental details}
A powder sample of TiSe$_2$ was synthesized via solid state reaction. Powdered Ti and Se were weighed out in the atomic ratio of Ti:Se 1:2.02 and ground together in accordance with Ref.~\cite{moya;prm19}. The resultant powder was sealed in a quartz tube under $\sim$3~Torr of partial argon atmosphere and heated at a rate of 50~$^{\circ}$C/hr to 650~$^{\circ}$C.  The sample remained at 650~$^{\circ}$C for 48 hours before being quenched to ambient temperature.

The powder sample of MnTe was the same one used in Ref.~\cite{baral;matter22}. It was prepared by thoroughly mixing stoichiometric amounts of Mn powder and Te pieces in an argon glove box, sealing the mixture in an evacuated quartz tube, and placing it in a furnace at 950~$^{\circ}$C for 6 hours. The ampoule was then quenched in ice water and annealed further at 650~$^{\circ}$C for 72 hours. A mortar and pestle were used to grind the sample into a fine powder in the glove box. 

The x-ray total scattering experiments were performed at the National Synchrotron Light Source II (NSLS-II) at Brookhaven National Laboratory on beamline 28-ID-1. The incident wavelength was 0.167~\AA. Finely ground powder samples were loaded into thin kapton capillaries sealed with clay. The capillaries were loaded into a liquid helium cryostat mounted on the beamline, allowing for continuous temperature control between 5 and 500~K, although the maximum temperature achieved in the present study was 300~K. A large amorphous silicon area detector was used to record the diffraction patterns, which were azimuthally integrated using DIOPTAS~\cite{presc;hpr15}. The integration was performed using 4096 radial points and 360 azimuthal points. The resulting one-dimensional diffraction patterns were normalized and Fourier transformed with $Q_{\mathrm{max}}=25$~\AA$^{-1}$ to produce the PDF using the xPDFsuite program~\cite{yang;arxiv15}.
 
The neutron total scattering experiments were performed at the Nanoscale-Ordered Material Diffractometer (NOMAD) at the Spallation Neutron Source (SNS) of Oak Ridge National Laboratory (ORNL)~\cite{neuef;nimb12}. The powder samples were loaded in quartz capillaries and mounted in the beam. Neutron scattering data were collected for an integrated proton current of 4~C and then reduced and transformed with $Q_{\mathrm{max}}= 25$~\AA$^{-1}$ using ADDIE~\cite{mcdon;aca17}, the automatic data reduction program at NOMAD.

\section{Results}
\subsection{TiSe$_2$}
The transition-metal dichalcogenide TiSe$_2$ is interesting for its complex electronic properties \cite{manze;nrm17, yin;csr21}, particularly the formation of a CDW below $\sim$200~K \cite{disal;prb76}. %Its ground state properties are also very sensitive to impurities and preparation conditions \cite{moya}. Symmetry mode driven PDF analysis was used to investigate local structural differences between samples prepared with varying cooling conditions at the end of a solid state preparation. \color{blue}[I'm not sure we want to go into the sample-dependent differences in this paper. I would keep the story simpler: 
A subtle structural distortion accompanies the CDW transition, making TiSe$_2$ a good test case for the ability to pick out small but long-range distortions in PDF data. This will be complementary to the MnTe test case, where the distortion is large but highly localized.

\begin{figure}
    \centering
    \includegraphics[width=12.0cm]{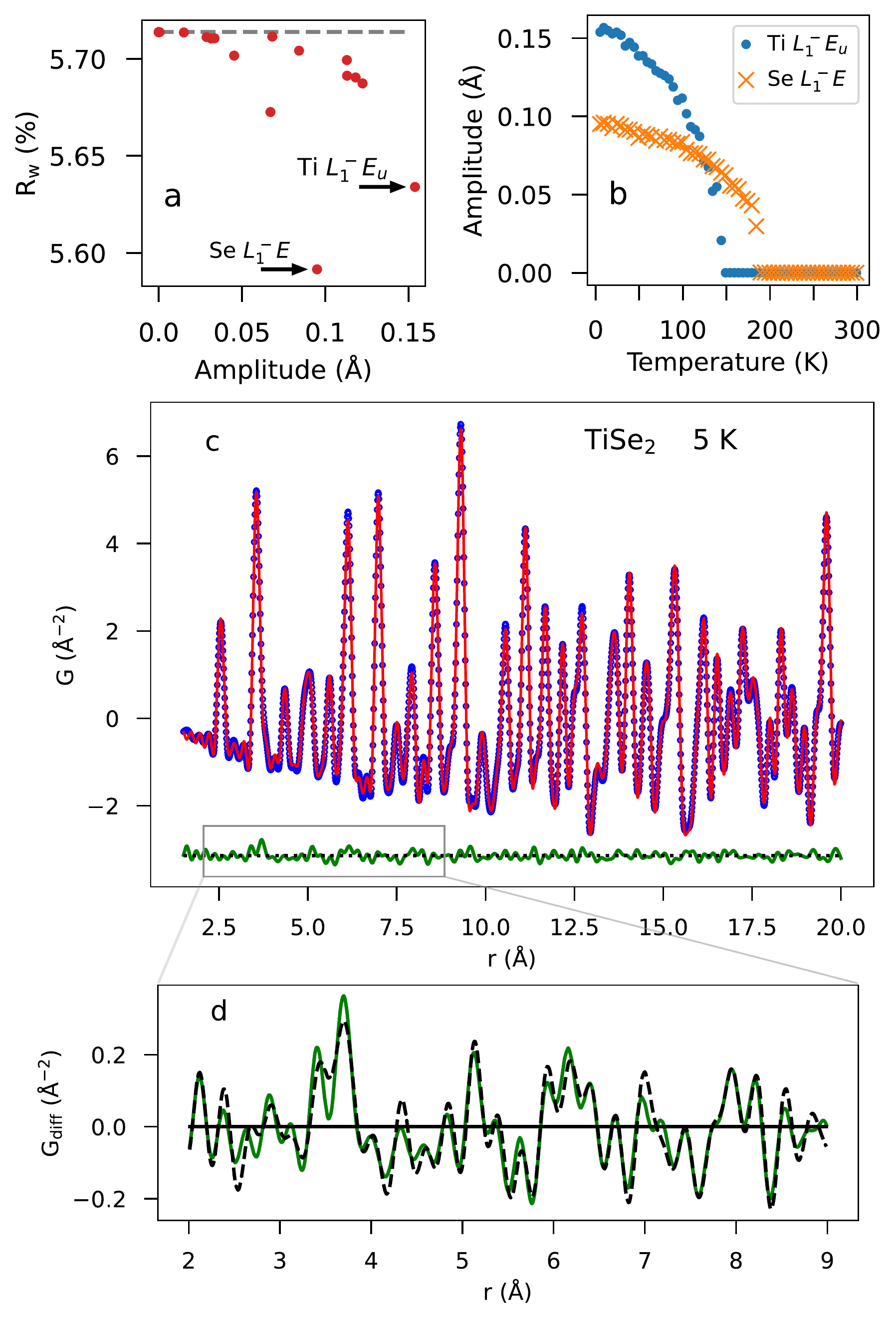}
    \caption{(a) A plot of \Rw\ versus mode amplitude for the 72 distortion modes. The arrows indicate the modes that improved the fit most significantly, namely the Ti $E_u$ and Se $E$ modes of the $L_1^-$ irrep. Note that 72 distinct points are not visible, because multiple modes may have the same effect on the calculated powder PDF pattern and therefore overlap on the plot. (b) Temperature dependence of the refined amplitudes of the Ti $E_u(a)$  and Se $E(a)$ modes of the $L_1^-$ irrep. (c) A representative PDF fit to data at 5.1~K with both modes active. The blue symbols represent the experimental data, the red curve the calculated PDF, and the green curve the fit residual, offset for clarity. (d) Comparison of a portion of the fit residual using the distorted model (green curve) and the undistorted parent model (dashed black curve).}
    \label{fig:TiSe_modes}
\end{figure}

The CDW state in TiSe$_2$ forms a $2\times2\times2$ superlattice of the $P\overline{3}m1$ high-temperature parent structure \cite{disal;prb76}. The supercell contains 24 atoms (8 formula units) and therefore a total of 72 degrees of freedom for $P1$ symmetry, which we selected to provide the most general possible exploration of potential distortions. For each of the 72 symmetry modes, we performed a PDF fit to the 5-K data from 1.5 - 20~\AA\ by first optimizing the lattice parameters and scale factor, then adding the symmetry-lowering mode amplitude and the displacement allowed in the parent space group ($P\overline{3}m1$), and finally adding the ADPs. To quantify the impact of each symmetry mode, we use the goodness-of-fit metric %$R_w = \sqrt{\left(\sum_i w_i(r_i) \left[G_{\mathrm{obs}}(r_i)-G_{\mathrm{calc}}(r_i) \right]^2 / \sum_i w_i(r_i) G_{\mathrm{obs}}^2(r_i)\right)}$
$R_w = \sqrt{\left(\sum_i \left[G_{\mathrm{obs}}(r_i)-G_{\mathrm{calc}}(r_i) \right]^2 / \sum_i  G_{\mathrm{obs}}^2(r_i)\right)}$, where $r_i$ is the $i^{\mathrm{th}}$ value in the $r$ grid of the experimental data, $G_{\mathrm{obs}}$ is the observed PDF, and $G_{\mathrm{calc}}$ is the calculated PDF. A lower value of \Rw\ indicates better agreement with the data. In Fig. \ref{fig:TiSe_modes}(a), we plot  \Rw\ versus mode amplitude for all 72 modes tested individually, along with a horizontal line at $R_w = 0.0571$ marking the base value with no symmetry-breaking modes included.
We see that most of the modes provide minimal improvement to the fit. However, two groups of modes improve the fit significantly more than the others, indicated by the arrows on the figure. These are the Ti $E_u$ and the Se $E$ modes of the $L_1^-$ irrep, corresponding to in-plane distortions of the Ti and Se atoms (refer to Fig.~\ref{fig:sym_mode}). Each group of modes includes three equivalent distortions that are rotated 120$^\circ$ from each other and, given the 1D nature of the PDF data, have identical effects on the calculated PDF. The $L_1^-$ irrep to which these modes belong is precisely the one responsible for the known CDW distortion in TiSe$_2$, which consists of equal superpositions of the three branches of the Ti $E_u$ and the Se $E$ modes~\cite{subed;prm22}. That the PDF fits naturally pick out these modes validates the use of this approach for TiSe$_2$ and confirms the ability of this symmetry-motivated fitting strategy to identify even rather subtle distortions. These same modes likewise provided the most improvement by a wide margin for longer-ranged fits up to 50~\AA. The fit was improved further when the same branches of the Ti and Se modes were refined together (e.g. $E_u(a)$ and $E(a)$ for Ti and Se, respectively), lowering \Rw\ from 0.0571 to 0.0542 for the 1.5 - 20~\AA\ x-ray fits. This difference is shown to be statistically significant in the Supplemental Information (SI).  The corresponding fit is displayed in Fig.~\ref{fig:TiSe_modes}(c). A portion of the fit residual is shown in Fig.~\ref{fig:TiSe_modes}(c), together with the fit residual corresponding to the undistorted structure for comparison. Although the difference is subtle, the distorted model is seen to systematically reduce the fit residual, excepting a few distinct points. The refined mode amplitudes at 5~K equate to displacements of 0.047(3) and 0.024(4)~\AA\ for Ti and Se, respectively, with the Ti-Se nearest-neighbor distance decreasing by approximately 0.05~\AA. These values are close to previously published results using traditional neutron and x-ray diffraction~\cite{disal;prb76,fang;prb17}.

%In testing combinations of these two mode groups, we found that only combinations of the same branch (i.e. Ti:$E_u(a)$ with Se:$E(a)$, Ti:$E_u(b)$ with Se:$E(b)$, and Ti:$E_u(c)$ with Se:$E(c)$) improved the fit significantly more than just a single distortion mode, and that none of the branches outperformed the others (to be expected, given that powder data cannot distinguish between the three equivalent directions of the distortion mode branches). Using ISODISTORT, we found that the symmetry with active Ti $E_u(a)$ and Se $E(a)$ modes corresponds to space group 15, $C2/c$. This is consistent with recent density functional theory calculations~\cite{subed;prm22}. 

Having identified the active distortion modes at 5~K, we then performed fits to the remaining x-ray PDF data sets with the Ti $E_u(a)$ and Se $E(a)$ modes included individually to examine the temperature dependence of the distortion. As illustrated in Fig.~\ref{fig:TiSe_modes}(b), we see that the mode amplitudes decrease with increasing temperature until reaching zero between approximately 150 and 200~K, corresponding reasonably well to the accepted CDW transition temperature around 200~K~\cite{disal;prb76}. The discrepancy in temperature may be due to the difficulty of resolving very small displacements near the transition in our data; this is consistent with the fact that the reduction in \Rw\ and the mode amplitude become nonzero at the same temperature (not shown in the figure). The figure demonstrates how the mode amplitudes serve as order parameters to identify structural transitions. Interestingly, the refined Se $E(a)$ mode amplitude becomes nonzero below 190~K, while the Ti $E_u(a)$ mode amplitude does so only below about 145~K. This may indicate that the Se distortion drives the CDW transition, but it may also occur simply because the weaker scattering strength of Ti reduces the sensitivity of the fits to Ti displacements compared to Se displacements. Additional studies could provide further clarity on this. Finally, we compare these results to fits performed to neutron PDF data collected at 100 and 300~K. The fits at 100~K yielded mode amplitudes of 0.18(3) and 0.08(2)~\AA\ for the Ti $E_u(a)$ and Se $E(a)$ modes, respectively. The corresponding values determined from the x-ray fits are 0.12(1) and 0.08(1)~\AA. At 300~K, the mode amplitudes refined to values below 0.0002~\AA, which is not meaningfully different from zero. This is expected in the undistorted state above the CDW transition.

\subsection{MnTe}
\begin{figure}
    \centering
    \includegraphics[width=10.0cm]{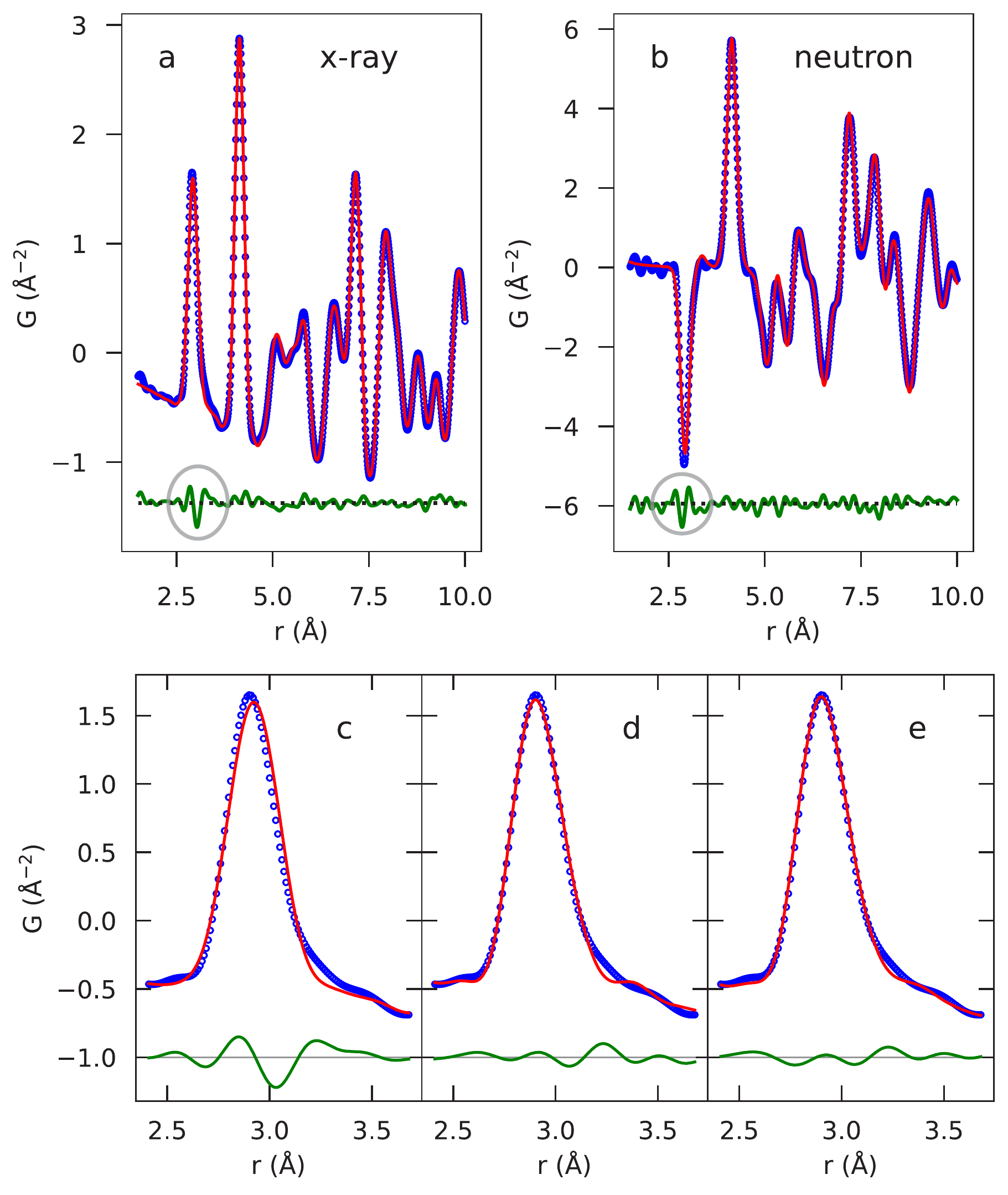}
    \caption{(a) X-ray PDF fit to MnTe at 300~K using the published hexagonal structure~\cite{dsa;jmmm05}. The blue symbols and red curve show the observed and calculated PDF, respectively, while the green curve shows the fit residual (offset vertically for clarity). (b) Same as (a), but for neutron PDF data (with the magnetic PDF~\cite{frand;aca14, frand;aca15} included in the fit). (c) Zoomed in view of the x-ray PDF fit in panel (a) highlighting the failure of the model for the first peak. (d) Fit when the Te $E'(a)$ mode of the $\Gamma_5^{+}$ irrep is active. (e) Fit when the Te $E'(a)$ mode of the $\Gamma_5^{+}$ irrep and the Mn $E_u(a)$ mode of the $\Gamma_6^{-}$ irrep are active. }
    \label{fig:MnTe}
\end{figure}

Hexagonal MnTe (space group $P6_3/mmc$) is an antiferromagnetic semiconductor that has garnered interest as a potential high-performance thermoelectric material~\cite{xu;jmchema17,ren;jmchemc17,dong;jmchemc18,zheng;sadv19,baral;matter22} and as a platform for antiferromagnetic spintronics~\cite{krieg;nc16}. In the context of the present work, MnTe is a useful example because PDF analysis reveals a large short-range distortion of the local structure that exists at room temperature. This is manifest as a significant misfit below approximately 3.5~\AA\ in the PDF fits when the published hexagonal structure is used~\cite{dsa;jmmm05}, as shown by the circled portion of the fit residual in Fig.~\ref{fig:MnTe}(a) and (b) for x-ray and neutron PDF, respectively. The data were collected at 300~K. A zoomed-in view of the misfit for the x-ray PDF pattern appears in Fig.~\ref{fig:MnTe}(c), revealing that the calculated peak is off-centered relative to the observed peak. The origin and significance of this feature in the local structure will be discussed elsewhere, but the objective here is to demonstrate that a symmetry-driven approach can be successfully applied to highly localized structural distortions, in addition to long-range distortions such as the example given in TiSe$_2$.

To investigate the local distortion in MnTe, we systematically tested each of the 12 displacive symmetry modes in the conventional unit cell assuming $P1$ symmetry. We restricted the fit range to 1.5 -- 3.75~\AA, since the misfit is localized to the first large peak in the PDF pattern. To avoid over-fitting to this short data range, we fixed the lattice parameters to the values determined from a longer-range fit (1.5 -- 20~\AA) and optimized only a scale factor, an isotropic thermal factor for each atomic species, and the symmetry mode amplitude. The peak-sharpening parameter $\delta_1$ was fixed to the value determined from the longer-range fit. 

\begin{figure}
    \centering
    \includegraphics[width=6.0cm]{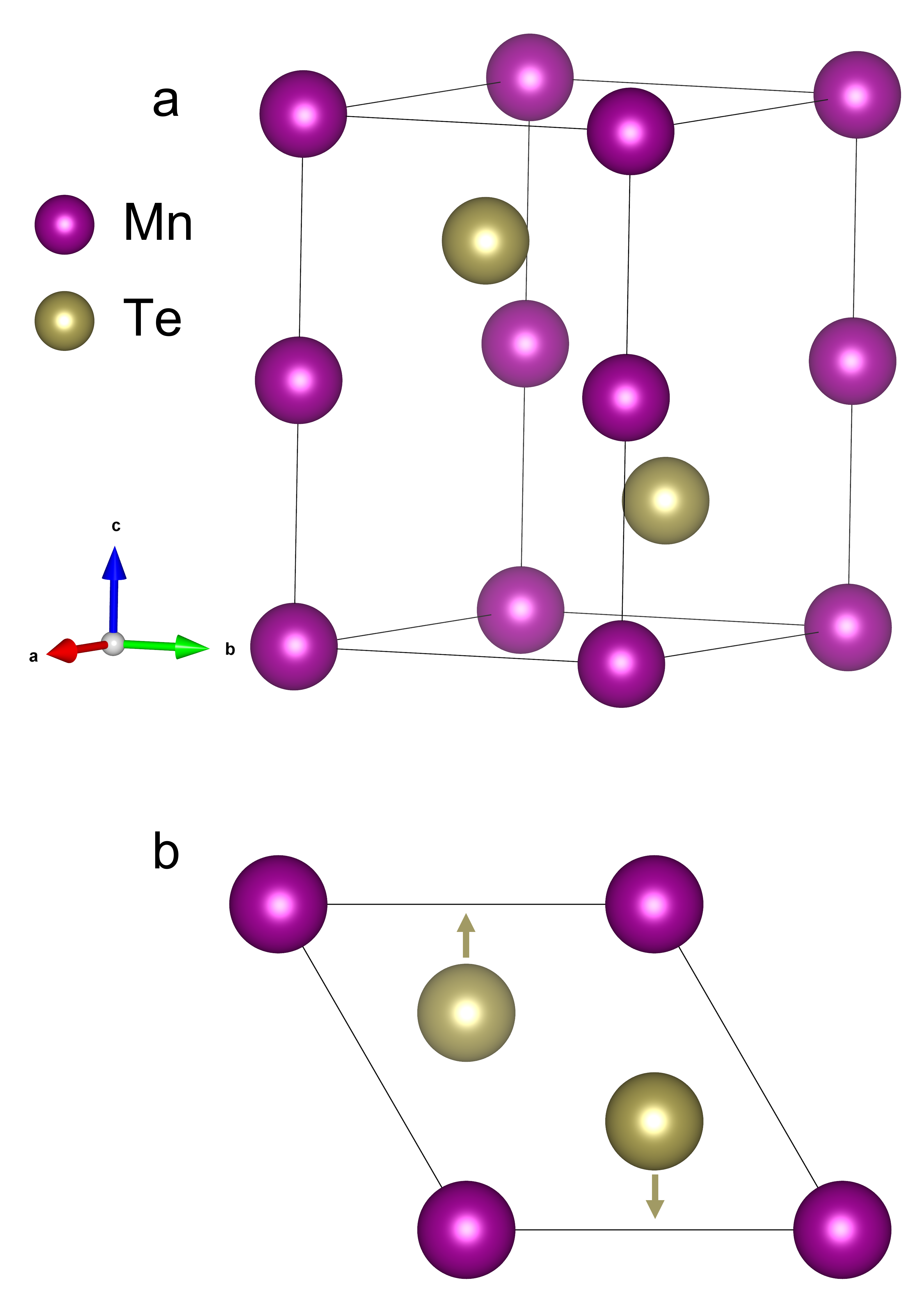}
    \caption{(a) Conventional unit cell of MnTe, visualized with VESTA~\cite{momma;jac11}. (b) View of the unit cell looking down the $c$ axis, with the Te displacements of the $\Gamma_5^{+} E'(a)$ shown by the arrows. }
    \label{fig:MnTe-distortion}
\end{figure}

Of the 12 symmetry mode amplitudes, 11 had only a small effect on the fit when tested individually. However, the Te $E'(a)$ mode of the $\Gamma_{5}^{+}$ irrep improved the fit dramatically, reducing $R_w$ from 0.115 to 0.058 for the x-ray fit and from 0.106 to 0.058 for the neutron fit. A plot of \Rw\ versus mode amplitude is shown in the (SI).  Both the neutron and x-ray data sets yielded a mode amplitude corresponding to a Te displacement of $\sim$0.126(4)~\AA. The improved fit is shown for the x-ray data in Fig.~\ref{fig:MnTe}(d); notably, the calculated peak is now centered on the experimental peak. Equivalent plots for the neutron PDF analysis are shown in the SI. The Te $E'(a)$ mode of the $\Gamma_{5}^{+}$ corresponds to antiparallel displacements of the two Te atoms toward the nearest face of the unit cell, as illustrated in Fig.~\ref{fig:MnTe-distortion}. If this were a coherent distortion throughout the entire crystal structure, the crystallographic symmetry would be lowered to space group $Cmcm$.
 %, splitting the six-fold degenerate nearest-neighbor Mn-Te peak of the undistorted structure into four short bonds ($\sim$2.873~\AA) and two long bonds ($\sim$3.024~\AA)
 
Refining pairs of symmetry mode amplitudes simultaneously improves the fit slightly. Fig.~\ref{fig:MnTe}(e), for example, shows the x-ray fit when the same Te $E'(a)$ mode of the $\Gamma_{5}^{+}$ irrep and the Mn $E_u(a)$ mode of the $\Gamma_{6}^{-}$ irrep are both active. The refined amplitudes were 0.09~\AA\ for $\Gamma_{5}^{+}$ and -0.17~\AA\ for $\Gamma_{6}^{-}$, and the value of \Rw\ was 0.038. However, other pairs of modes yielded equally good fits, indicating that the data range over which the local distortion is observed is insufficient to remove all ambiguity. Whether or not the peak-sharpening parameters $\delta_1$ or $\delta_2$ were included also influenced which mode pairs yielded the best fit. On the other hand, the single-mode refinements were robust against variations in the choice of $\delta_1$ or $\delta_2$ (or no peak-sharpening function at all), the number of unique ADPs, and the starting values of the variables, building confidence in the reliability of the single-mode results. Finally, we note that symmetry-mode fits performed against a data range exceeding $\sim5$~\AA\ were unable to fit the low-$r$ and high-$r$ peaks simultaneously; instead, the fits yielded negligibly small mode amplitudes such that the longer-range peaks were well described at the expense of the first large peak. This indicates that the distortion in MnTe is highly localized.

\section{Discussion and Conclusion}
We demonstrated the utility of the symmetry-mode approach to fitting PDF data for two complementary use cases. In the case of TiSe$_2$, we identified a structural phase transition present over the full $r$ range of the data in a more straightforward manner than would have been possible using traditional PDF fitting methods with $xyz$ atomic coordinates as fit variables. The amplitudes of the active symmetry modes served as structural order parameters to show subtle but observable changes to the long-range structure, consistent with published results. Here, the advantage of the symmetry-driven methods over the traditional $xyz$ approach is the reduced number of free parameters necessary to test and identify distorted structural models that fit the data. Only two nonzero mode amplitudes were necessary to characterize the distorted structure, while systematic testing of the individual atomic coordinates would have required a much larger number of fit variables to arrive at the same result.

The MnTe case offers a proof of concept for fitting distortions with shorter correlation lengths, as well. The distortion was only apparent in the data below about 4~\AA. The flexibility of our implementation allows for the isolation, or concurrent testing, of individual modes and groups of modes as defined by the user. For MnTe, we identified a single Te mode that resulted in a much better fit when activated. The extremely localized nature of this distortion obscures the notion of symmetry modes somewhat, but the symmetry mode basis nevertheless provided a useful parameterization, leading us to identify an in-plane distortion of the Te atoms as the likely cause of the misfit in the original fit. 

A benefit of using the flexible DiffPy fitting framework is that users can customize the fitting procedure in any way they desire. For example, one could easily execute the fit multiple times with random starting values, impose a penalty to the cost function for each active symmetry mode, or set a threshold improvement value in \Rw\ to allow a mode to be considered active. These strategies have all been suggested previously as a way to improve the reliability of exhaustive symmetry-mode testing~\cite{bird;jac21, kerma;aca12}. Our code can also be extended in a straightforward manner to include the rotational, occupational, and magnetic modes calculated by ISODISTORT, although that is beyond the scope of the present work. The study of molecular materials can also benefit from symmetry-mode analysis if the molecules contain rigid units~\cite{liu;jacs18}. We expect that symmetry-adapted PDF analysis will become increasingly common in the field of total scattering, leading to new physical insights into the local structure of materials.

\section{Code availability and usage}
The analysis methods introduced in this work are based on fully open-source software. We have developed two new packages as part of this work: \texttt{isopydistort} (\url{https://github.com/FrandsenGroup/isopydistort}) for automated interactions with the ISODISTORT web-based software and \texttt{isopytools} (\url{https://github.com/FrandsenGroup/isopytools}) for adapting the ISODISTORT output to be compatible with the \texttt{diffpy} library~\cite{juhas;aca15}. Example scripts and data files corresponding to the analysis presented in this work are also available in the Supplemental Materials.

     % Appendices appear after the main body of the text. They are prefixed by
     % a single \appendix declaration, and are then structured just like the
     % body text.

\ack{Acknowledgements}
We thank Branton Campbell for valuable discussions regarding symmetry-adapted distortion modes and ISODISTORT. We acknowledge Brian Toby and Robert von Dreele, who developed the first python code to interface with ISODISTORT as part of their GSAS-II program. Work by P.K.H., R.B., and B.A.F. was supported by the U.S. Department of Energy, Office of Science, Office of Basic Energy Sciences (DOE-BES) through Award No. DE-SC0021134. A.M.H. acknowledges supper from the Natural Sciences and Engineering Research Council of Canada (NSERC) and the CIFAR Azrieli Global Scholars program. J.M.M. was supported by the National Science Foundation Graduate Research Fellowship under Grant DGE 1842494. This study used resources at the Spallation Neutron Source (SNS), a DOE Office of Science User Facility operated by the Oak Ridge National Laboratory. This research used beamline 28-ID-1 of the National Synchrotron Light Source II, a U.S. Department of Energy (DOE) Office of Science User Facility operated for the DOE Office of Science by Brookhaven National Laboratory under Contract No. DE-SC0012704.

\end{document}